\documentclass[12pt,aps,nofootinbib]{revtex4}
\usepackage{bm,graphicx,amsmath,amssymb}%
\usepackage[mathscr]{euscript}
\newcommand*{\dis}{\ell}
\newcommand*{\ogr}[2]{#1\,\vrule \,{}_{\displaystyle{}_{#2}}}
\newcommand*{\rmd}{\mathrm{d}}
\newcommand*{\ssy}[5]{#1, \emph{#2} {\bf #3}, #5 (#4)\rlap{.}}
\newcommand*{\av}[1]{\ensuremath{\langle\,{#1}\rangle_0}}
\newcommand{\sred}{\bar{\varrho}}
\newcommand{\karti}[4]{\begin{figure}[#1]\begin{center}
\includegraphics[height=#2]{#3}
\end{center}\caption{#4}\end{figure}}
\newcommand{\px}[1]{\usefont{U}{pxsyb}{m}{n}\selectfont
#1}
\newcommand{\rea}{\mathord{\mbox{\px R}}}
\newcommand{\mink}{\mathord{\mbox{\px L}}}
\newcommand{\tot}{\ensuremath{E_{\rm tot}^-}}
\begin{document}
\title{The quantum inequalities do not forbid spacetime shortcuts}
\author{S. Krasnikov\thanks{Email: redish@gao.spb.ru}}
\affiliation{The Central Astronomical Observatory at Pulkovo}
\date{}

\begin{abstract}
A class of spacetimes (comprising the Alcubierre bubble, Krasnikov
tube, and a certain type of wormholes) is considered that admits
`superluminal travel' in a strictly defined sense. Such spacetimes
(they are called `shortcuts' in this paper) were suspected to be
impossible because calculations based on `quantum inequalities'
suggest that their existence would involve Planck-scale energy
densities and hence unphysically large values of the `total amount of
negative energy'
\tot.
\par I argue that the  spacetimes of this type
may not be unphysical at all. By explicit examples I prove that: 1)
the relevant quantum inequality does not (always)  imply large energy
densities; 2) large densities may not lead to large values of
\tot; 3) large \tot\, being physically meaningless  in some relevant
situations, does not necessarily exclude  shortcuts.
\end{abstract}
\maketitle
\section{Introduction}
Suppose the distance from the Earth to a star, found by usual
astronomical methods (by measuring the parallax, say), is 100 light
years. Suppose also that (in agreement with all that we know) no body can
move faster than light. It is then tempting to conclude that a
spaceship sent to that star cannot return sooner than in 200$\,$yr.
However, such a conclusion
may be too hasty. The point is that the quantity $D_p$, which defines the
travel time, and the quantity $D_a$ measured by the parallax, in general
relativity (in contrast with special relativity) are not the same, even
though, duly defined, they both deserve the name `distance'. In
practice, some approximately flat (and pretty narrow) region $\mathcal
R$ is considered, comprising the Earth,  the star, and a geodesic
connecting them. $D_a$ then is defined by that geodesic \emph{as if}
$\mathcal R$ were a part of the Minkowski space. That the thus defined
$D_a$ may be much greater than $D_p$ can be seen already from the fact
that beyond $\mathcal R$ a short wormhole  may occur, which connects
the vicinities of the Earth and of the star (as we shall see in a
moment a non-trivial topology is \emph{not} essential for the matter
in discussion). A spaceship then can take a short cut through the
wormhole and thus make the trip \emph{faster than light}. Of course
the words `faster than light', used in such a context do not mean that
the spaceship locally (when the notion of speed is well defined) moves
faster than a passing photon. Actually in the above example we
compared two different spacetimes --- the real world with a wormhole
and a \emph{fictitious} Minkowski space by which we \emph{erroneously}
described our world --- and found that the travel time of a spaceship
in the former is less than that of a photon in the latter.

Generalizing the above example (in the next section we briefly discuss
two other possible ways to give a precise meaning to the words
`faster-than-light' in application to  non-tachyonic objects) we
introduce the following notion.
\paragraph*{Definition.} Let $C$ be a timelike cylinder
$\sum_{i=1}^3x_i^2\leqslant c^2$
in the Minkowski space $\mink^4$. A globally hyperbolic
spacetime $M$ is a \emph{shortcut} if there are a region $U\subset M$,
an isometry $\varkappa\colon\; (\mink^4-C)\to U$, and a pair of points
$p,q\in \mink^4-C$ such that
$$
p\not\preccurlyeq q,\qquad \varkappa(p)\preccurlyeq\varkappa(q).
$$
\par\noindent
In other words we call $M$ a shortcut if it can be obtained from the
Minkowski space $\mink^4$ by replacing a flat cylinder
$C\subset\mink^4$  with something else so that some spacelike
separated (in $\mink^4$) points are causally connected in $M$.
\paragraph*{Examples.}
 Consider a  plane $\rea^2$
with the metric
$$
\rmd s^2 = \rmd r^2 + R^2(r)\,\rmd\phi^2,
$$
where
$$
r\geqslant D_a-d,\quad \phi=\phi+2\pi,\qquad \ogr{R}{r<D_a}=r-D_a+d,
\quad \ogr{R}{r>D_a+\delta}=r.
$$
The plane is flat except in a thin annulus $D_a<r<D_a+\delta$. However, an
observer at $r=D_a+\delta,$ $\phi=0$ is much closer --- $2(d+\delta)$
against $2(D_a+\delta)$ --- to the diametrically opposite
point $r=D_a+\delta,$ $\phi=\pi$ than  if the
\emph{whole} plane were flat. A
four-dimensional generalization of such a spacetime
\begin{equation}\label{eq:fdige}
\rmd s^2 =\rmd t^2 - \rmd r^2 - R^2(r)(\rmd
\theta^2+\sin^2\theta\rmd\phi^2)
\end{equation}
is shown in
figure~\ref{fig:shortcut}a.
\karti{t,b}{0.3\textwidth}{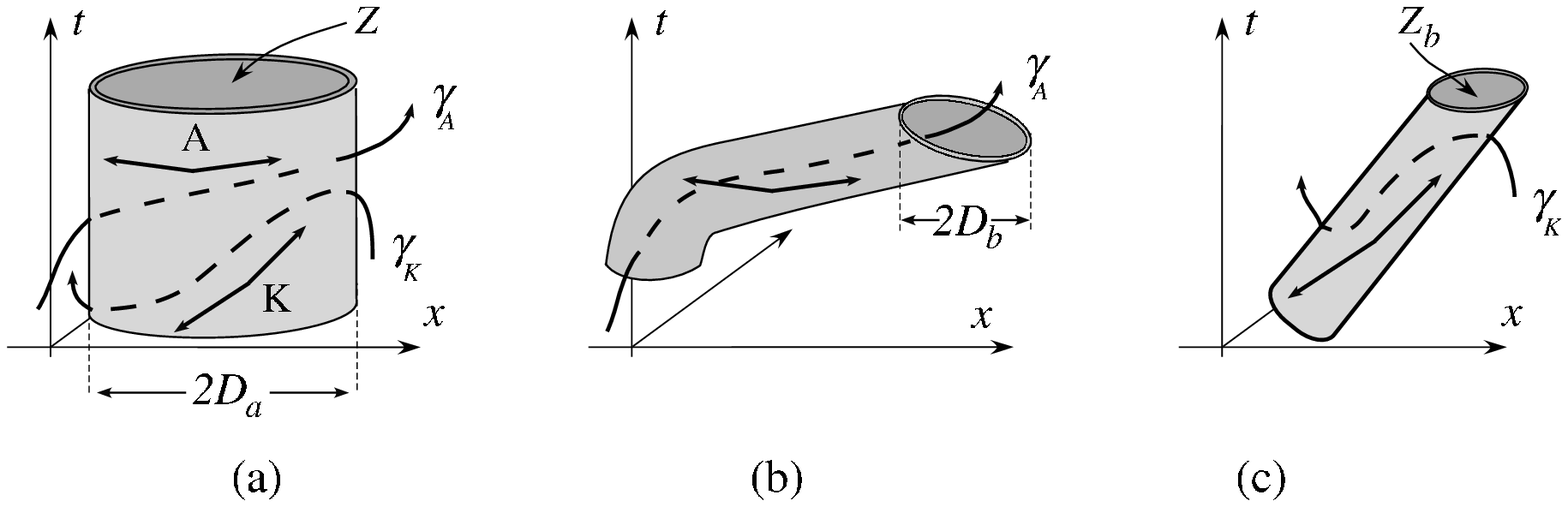}{\label{fig:shortcut} The
domains bounded by the gray walls may be flat and, nevertheless,
differ significantly from $C$. The pairs A and K of null vectors
correspond to Alcubierre and Krasnikov metrics,
respectively.}%
The metric inside the cylinder $Z$ is as flat as the metric outside,
but the null cones are `more open' (this of course is a coordinate
effect and would not take place, say, in the spherical coordinates
\eqref{eq:fdige}) and thus curves like $\gamma_A$ are timelike, though
they would be spacelike, if the metric were Minkowski in the whole
spacetime. To use the described phenomenon for interstellar travel one
need not create for a single trip such a huge cylinder with $D_a$ of the
order of light years. It would suffice to surround the pilot with a
small bubble --- called the \emph{Alcubierre bubble}\footnote{The
metric originally proposed by Alcubierre \cite{Alc} is a little
different (in particular, the null cones also tilt inside the bubble,
not just open), but the principle of operation is the same.} --- with
the diameter $D_b\ll D_a$ (see figure~\ref{fig:shortcut}b), which would
reduce the area of the `domain wall' surrounding the pilot --- or,
rather, of its outer surface --- by $10^{32}$ (for $D_a\sim 100\,$ly and
$D_b\sim 100\,$m).

The null cones in $Z$ (or in $Z_b$ see figure~\ref{fig:shortcut}c) can
be tilted so much (with the metric remaining flat) that a
\emph{future-directed} null vector is directed in the sense of
decreasing $t$ \cite{FTL} (which, being again a coordinate effect, has
no direct connection with causality --- the spacetime is globally
hyperbolic). The difference between such a shortcut (called the
\emph{Krasnikov tube} \cite{EveRo}) and the Alcubierre bubble is of no
significance for problems discussed in this paper [it becomes crucial
in situations (see below), when one have to consider \emph{round} trips].
\paragraph*{Remark.} Creation of a shortcut in itself requires some
time. Moreover, under some mild assumptions about the properties of
the `building materials' (the absence of tachyons, essentially) this
time is \emph{too} large \cite{FTL}  in the following sense. If a
$D_a=100$ light years, then a shortcut that would allow one to reach
the star in 1$\,$yr, cannot be built in less than 99$\,$yr. So, a
\emph{created, artificial} shortcut (in contrast to a \emph{found, natural}
one, such as a relic wormhole, say) is a useful means of interstellar
travel only when the large time taken by the first trip is not
important \cite{FTL,Coule}. This may be an exploratory expedition,
(when only the time of return matters), or  a regular
space service between two stars.

It seems interesting from both academic and practical points of view
to find out whether something like a shortcut can be found in
nature or manufactured  by an advanced civilization. Referring
to an advanced civilization I mean that we are not concerned with
`technical details' such as: how to create a wormhole, how to
penetrate a domain wall, etc. At this stage of research almost any
spacetime may be acknowledged as being possible unless it clearly
contradicts some fundamental laws or observations. It appears,
however, that even by such  liberal criteria the existence of
shortcuts is questionable. As was shown in \cite{EveRo} maintaining an
Alcubierre bubble 100$\,$m in diameter involves energies of the order
of $\sim 10^{67}\,\text{g}\approx 10^{34}M_\odot$. A similar result
was obtained in \cite{Pfen97b} for the Krasnikov tube and, as we argue
below, can be, analogously obtained for a traversable wormhole as
well. Such a figure looks absolutely discouraging even with regard to
an extremely advanced civilization and can be viewed as a prohibition
of shortcuts.
\par
The goal of this paper is to show that it is not impossible to get around
this prohibition. This will be shown in section~\ref{sec:ways} after a
brief discussion --- in section~\ref{sec:restr} --- of the origin and
meaning of that awesome figure.
\section{Restrictions on shortcuts}
\label{sec:restr}
\subsection{Superluminal travel and the weak energy condition}
\label{subs:wec} The root of the problems involved in creating a shortcut
lies in the fact that,  as was shown in
\cite{Tho}, \cite{Alc}, and \cite{EveRo} for wormholes, Alcubierre's
bubble,
and Krasnikov's tube, respectively, the Weak energy condition (WEC)
$$
G_{\mu\nu}t^\mu t^\nu\geqslant 0\qquad \forall \ \text{timelike}\ \bm t
$$
must break down in some regions of these spacetimes. When the
\emph{classical} Einstein equations are adopted the violation of WEC implies
that for some observers the energy density $T_{00}$ in those regions is
negative, which is forbidden.
\par
That a shortcut requires negative energy is hardly surprising --- if
the energy density were non-negative everywhere one would expect  the
total mass of the shortcut to be positive, while in fact it is zero by
definition.  Hence an important question  arises:
Is the necessity of negative energy densities something
inherent in superluminal travel, or is it just an artifact of our
approach, in which the spacetime outside some region is required to be
\emph{strictly} flat\footnote{Note in this connection that,
unfortunately, it is hard (if possible at all) to define reasonably a
`nearly flat' spacetime. This is because, in contrast to the
Riemannian case, there is no set of scalars such that their vanishing would
guarantee that a spacetime is flat \cite{wave}.} and thus the gravitational
fields of stars, nebulae, and other potential
sources of the energy needed for the trip are neglected.
It is instructive to compare the notion of a shortcut with its
alternatives.

Consider a group of runners. The starting line and the finish line
taken, respectively, at the moments, when the race begins and when the
first runner finishes, are spacelike geodesics. We recognize a runner
as the fastest if he or she is the only one whose world line
intersects both these geodesics. Olum proposed \cite{Olum} to use the
same criterion to distinguish a `fastest' null geodesic, which he
calls superluminal (see also \cite{gaojul}). More specifically a null
geodesic $\gamma$ connecting points $p$ and $q$ is called superluminal
(in the (2+1)-dimensional case, which is easier to visualize) only if
there are geodesics $\lambda_p\ni p$ and $\lambda_q\ni q$ such that of
all their points only $p$ and $q$ are causally connected. And it is
proved in \cite{Olum} that if the generic condition holds on $\gamma$,
then WEC does not.

The advantage of Olum's definition (or, rather, part of the definition,
since only a necessary condition is formulated) is that one does not need
compare objects and quantities belonging to different spacetimes. On
the other hand, it excludes some paths which (on the same intuitive
grounds) one might want to consider as superluminal. Suppose, for
example, that $\lambda_p$ is the line $t=0$, $x=-2c$ in a shortcut $M$
(the coordinates are pulled back by $\varkappa$ from the Minkowski
space) and that the photons emitted from this line move with a constant
$y$ and meet the plane $x=2c$ in points $t=f(y)$. Now, according to
Olum, if $f$ has a
\emph{strict} minimum (in $y_0$, say),  it   qualifies the geodesic
emitted in $\lambda_p(y_0)$ as being superluminal, but not if the
minimum is weak (which, for example, is the case with a `portal' (see
subsection~\ref{subs:portal}), or an Alcubierre bubble if its front
wall is flatten).

A different approach is proposed in \cite{FTL} (see also \cite{Low}).
The travel time $T_{\rm R}$ in the real world $M_{\rm R}$ --- with all its
shortcuts (if they are present there) and with (possibly non-trivial)
interaction between the spaceship and its environment --- is again
compared with the travel time $T_{\rm F}$ in some fictitious spacetime
$M_{\rm F}$. This time, however, $M_{\rm F}$ is not the Minkowski space, but the
spacetime that \emph{would have} formed if the trip had not been
undertaken. As an example (for the rigorous definition and some
discussion see \cite{FTL})) consider a spherically symmetric spacetime
which is empty outside a cylinder $N=B_{r_o}\times\mink^1$, where
$B_{r_o}$ is a (3-dimensional) ball with the radius $r_o$. Let us
interpret this spacetime as a model of a neighbourhood of a globular
cluster. At $t=0$ an observer located in a point with
 $r=r_S>r_o$ decides to explore a star located in the
diametrically opposite point and considers the following two
scenarios. First, he/she can send a photon, which would pass through
the cluster, reflect (at $t=T_D$, say) from something near the desired
star, pass through the cluster once again (in the opposite direction),
and, finally,  return to $r=r_S$. The other possibility is to send a
spaceship instead of the photon. Being powerful enough, the spaceship
on its first passage through the cluster would push the stars, blow
them up, emit different (but all satisfying WEC) fields, etc. As a
result the metric at $t=T_D$ in this second scenario differs from that
in the first (though, if the spherical symmetry is preserved by the
pilot, it may remain, say, Schwarzschild, outside $N$). So, it is not
surprising that the back way --- and thus the whole trip --- takes
different time ($T_{\rm F}$ and $T_{\rm R}$, respectively) for the
photon in the first scenario  and for the spaceship in the second. And
if $T_{\rm R}<T_{\rm F}$ such a trip deserves to be called
superluminal. A superluminal, in this sense, journey does not require
violation of WEC. In the appendix we prove this fact (though actually
it is  almost self-evident) by constructing a specific example.
\par
The problem with this example is that $T_{\rm R}$ in it, though being
less than $T_{\rm F}$, is still greater than $T_{\rm M}$, where
$T_{\rm M}=4r_S$ is the time required for the same trip in the
Minkowski space (by the `same' trip I mean the trip between the
points with the same $t,r,\phi$, and $\theta$ coordinates; due to the
spherical symmetry and staticity of $M_{\rm F}$ and $M_{\rm R}-N$ such
a mapping on the Minkowski space is more or less meaningful, see
\cite{gaojul} though). So one can interpret $T_{\rm F}-T_{\rm R}$ to
be not so much a gain in travel time as some compensation of the time
delay $T_{\rm F}-T_{\rm M}$ experienced by a traveler in $M_{\rm F}$
and caused by the variation of $M_{\rm F}$ from the Minkowski space.

\subsection{The `total negative energy' $\bm{\tot}$}\label{subsec:tot}
Since the classical matter (alone) cannot sustain a shortcut we turn our
attention to the quantum effects --- described within the
semiclassical approximation \cite{BirDav} --- and, correspondingly,
cast the Einstein equations into the form
\begin{equation}\label{eq:Ein}
G_{\mu\nu}=8\pi T^{\rm C}_{\mu\nu} + 8\pi \langle\, T_{\mu\nu}\rangle.
\end{equation}
Here $T^{\rm C}_{\mu\nu}$ is the contribution of the `classical
matter', that is the matter for which quantum effects can be
neglected. $T^{\rm C}_{\mu\nu}$ is supposed to obey the Weak energy
condition, but is otherwise arbitrary. As for the second term, it is
the (renormalized) expectation value of the stress-energy tensor of
quantum fields involved. As is well known $\langle\,
T_{\mu\nu}\rangle$ may violate WEC and so the necessity of negative
energy densities does not by itself exclude the shortcuts. It was
found, however (see
\cite{thes} for a review), that such violations are not arbitrary, but
may be subject to a restriction called the quantum inequality
(QI). The remainder of this subsection is a (very sketchy) review of
how QI in its turn imposes restrictions on shortcuts.
\par
Consider  the free electro-magnetic or scalar (massless,
minimally-coupled) field in the Minkowski space. Let $\sred_\chi$
be its energy density averaged with a weighting function $\chi$ over a
timelike geodesic $\gamma(\tau)$
\begin{equation*}
\sred_\chi(\tau_0,\Delta)\equiv\int_{-\infty}^{\infty}\,
\langle\, T_{\mu\nu}u^\mu u^\nu\rangle\chi(\tau_0,\Delta;\tau)\, \rmd \tau
\end{equation*}
($\tau$ is the proper time on
$\gamma$ and $\bm u\equiv\partial_\tau$). It is assumed that
$\chi(\tau)$ is smooth, its integral is unity, and its support lies in
$(\tau_0-\Delta/2,\tau_0+\Delta/2)$, so that, when $\Delta$ is small,
\begin{equation}\label{d2ro}
\sred_\chi(\tau_0,\Delta)\approx\varrho(\tau_0),
\end{equation}
where  $\varrho\equiv\langle T_{\mu\nu}u^\mu u^\nu\rangle$ is the
energy density of the field as measured by an observer with the world
line $\gamma$. This equality allows one to
 estimate the energy density, because $\sred_\chi$ obeys the following
 \emph{`quantum inequality'} \cite{ref:QI}
\begin{equation}\label{eq:QI}
\sred_\chi
 \geqslant -A\Delta^{-4},
\end{equation}
where $A$ is a positive constant of order of unity (from now on I
freely omit insignificant constants like $A$, $8\pi$, etc.).
\par
Assume now  (cf.~\cite{EveRo,Pfen97b}) that \eqref{eq:QI} is valid
also in \emph{curved} spacetime if $\Delta$ is sufficiently small, or,
more specifically \cite{Qicurv}, if the following  holds
(actually, one more inequality is implied, see the next section)
\begin{equation}\label{eq:supp}
 \Delta\lesssim \dis\equiv
 \big(\max |R_{\alpha\beta\gamma\delta}|\big)^{-1/2},
\end{equation}
where the components of the Riemann tensor are found in the observer's
proper frame. The condition \eqref{eq:supp} is supposed to guarantee
that the relevant segment of $\gamma$ lies in a region so small that
it can be regarded as `approximately Minkowskian'. The quantum
inequality \eqref{eq:QI} [allowing for \eqref{d2ro} and
\eqref{eq:supp}] relates the curvature in a point with the energy
density in this point. It restricts, loosely speaking,
the amount of `exotic matter' (i.~e.\ matter violating the Weak
energy condition \cite{Tho}) that may be produced by the curvature of
spacetime. But (almost) the same quantities are related also by the
Einstein equations $G_{00}=  8\pi\varrho$. Ignoring in our rough
consideration  the possible difference between the scales given by the
components of the Einstein and of the Riemann tensors, we can combine
the two relations to obtain
\begin{equation}\label{eq:chain}
\varrho
\sim\sred_\chi\geqslant -\Delta^{-4}\sim
-\dis^{-4}\sim- \varrho^2,
\end{equation}
which, when $\varrho$ is negative, gives
\begin{equation}\label{eq:est}
| \varrho|\gtrsim 1.
\end{equation}
It is this Planck-scale energy density that gives rise to the
prohibitive figures cited in the introduction (recall that $1\approx
5\times10^{93}\,$g/cm$^3$). Indeed, let us estimate the total amount
of negative energy \tot\ required for maintenance of a shortcut. We
define it as follows
\begin{equation}\label{eq:etot}
\tot\equiv \int_{\Xi} |\varrho |\,\rmd^3 x.
\end{equation}
Here we have chosen a spacelike surface $\mathcal S$ in $M$ and
denoted by $\Xi\subset\mathcal S$ the region within it in which WEC is
violated.
\par
\tot\ is approximately equal to $-\varrho
V_\Xi$, where $V_\Xi$ is the volume of   $\Xi$. In both Alcubierre and
Krasnikov spaces $\Xi$ is a spherical layer (`domain wall')
surrounding the domain  $\mathcal D$ of the `false' flat metric.
 The volume of the domain wall can be
estimated as $V_\Xi\gtrsim  S_{\:\rm i}\delta$, where  $S_{\:\rm i}$ is
the area of its inner surface (recall that as discussed in the
introduction, the area of the outer surface is much greater, even
though $\delta$ is small). $\mathcal D$ must be at least $\sim
1\,$m in diameter --- which means that $S_{\:\rm i}\gtrsim
 10^{70}$ --- to accommodate a human. For a spherically symmetric
wormhole, $\Xi$ is essentially the throat of the wormhole, that is
also a spherical layer with the radius $\gtrsim 1\,$m if a  human
being is supposed to pass through it. So, even if the thickness of the
layer $\delta\sim l_{\rm Pl}$, one might conclude that it would take
at least
\begin{equation}\label{eq:plotn}
\tot\sim\mathcal |\varrho S_{\:\rm i}\delta|\approx
10^{32}M_\odot
\end{equation}
of exotic matter to support a shortcut. Such a huge value presumably
indicates the `unphysical nature' \cite{Pfen97b} of shortcuts.
\paragraph*{Example.}
Consider a  Morris-Thorne wormhole
\begin{equation}\label{eq:r(l)}
  \rmd s^2=-\rmd t^2+ \rmd l^2
           +r(l)^2(\rmd\theta^2+ \sin^2\theta\,\rmd\phi^2),
\end{equation}
where $r(l)$ is a smooth even function with a single minimum
$r(0)=r_0$. At positive $l$ (i.~e.\ at $r>r_0$) we can choose $r$ to
be a coordinate and rewrite the metric in the following form
\begin{equation}\label{eq:mtwh}
  \rmd s^2=-\rmd t^2+\frac{r}{r-b(r)}\,\rmd r^2
           +r^2(\rmd\theta^2+ \sin^2\theta\,\rmd\phi^2),
\end{equation}
where $b(r)$ can be, if desired, expressed in terms of $r'(l)$. In the
special case in which\footnote{It is understood that actually $b(r)$ is
smoothed near $r=r_0+\delta$ in a manner that does not affect the
following considerations.\label{fo:smo}}
\begin{equation}\label{eq:absben}
   b(r)=
   \begin{cases}
   r_0[1-{(r-r_0)}/{\delta}]^2,&
   \text{at }r\in( r_0, r_0+\delta]\\
    0&
   \text{at } r\geqslant r_0 +\delta
   \end{cases},\qquad \delta \ll r_0
\end{equation}
the wormhole (\ref{eq:mtwh},\ref{eq:absben}) is
called \emph{`absurdly benign'} \cite{Tho}. The spacetime is flat
except for a spherical $\delta$ thick layer $\Xi$ in which $b\neq 0$. The
energy density in this layer is
$$
G_{\hat{t}\hat{t}}\sim -(\delta r_0)^{-1}
$$
and it is easy to see that (cf. endnote 25 of \cite{EveRo})
\begin{equation*}%\label{eq:frug}
\tot\sim G_{\hat{t}\hat{t}} V_\Xi\sim  r_0
\approx10^{-3}M_\odot\left(\frac{r_0 }{1\,\text{m}}\right),
\end{equation*}
which looks of course far more attractive than \eqref{eq:plotn}. Let
us see, however, what restrictions on the parameters of the wormhole
are imposed by QI. The maximal component of the  Riemann tensor in an
orthonormal frame of a static observer located in the throat (i.~e.\
near $r=r_0$) is
$$
\left| R_{\hat{r}\hat{\phi}\hat{r}\hat{\phi}}\right|
\approx \frac{1}{\delta r_0}
$$
Correspondingly, QI (\ref{eq:QI},\ref{eq:supp}) requires that
$$
\sred_\chi \gtrsim - (\delta r_0)^{-2}
$$
(as we shall see in a moment, this is a \emph{vastly} more restrictive
condition than $\sred_\chi \gtrsim - \delta^{-4}$, which is considered
in \cite{Qicurv}). $\varrho$ does not depend on $\tau$ for a so chosen
observer and hence
$$
\sred_\chi=\varrho=\ogr{G_{\hat{t}\hat{t}}}{r=r_0}
\sim -(\delta r_0)^{-1}
$$
Comparing this with the preceding inequality we find that
$\varrho\lesssim -1$, as expected, \emph{and} $\delta\lesssim 1/r_0$.
So, the frugality of the wormhole owes just to the fact that the
thickness $\delta$ of the curved region  is not of the order of
$l_{\rm Pl}$ (as we took in deriving
\eqref{eq:plotn}), but of the order of $10^{-35}l_{\rm Pl}$, which
is a value, of course, that makes the wormhole much more absurd than benign.

\section{Ways out}\label{sec:ways}
The analysis of the previous section leading to the restrictions
\eqref{eq:est} and \eqref{eq:plotn} and to their interpretation is quite
rough, of course. To some extent it can be refined, see
\cite{EveRo,Pfen97b,Qicurv}, but, as we discuss in this section,
significant loopholes remain.

In the search for realistic shortcuts an obvious line of attack would
be to look for situations in which  the quantum inequality \eqref{eq:QI}
does not hold. Note in this connection that for non-Minkowskian
spacetimes \eqref{eq:QI} has never been proved\footnote{The only
exception, to my knowledge, is the two-dimensional conformally trivial
case \cite{conf2}. There are also `difference inequalities' (see,
e.~g., \cite{FE}), but , though resembling \eqref{eq:QI} in
appearance, they differ from it fundamentally (they restrict a \emph{part}
$\langle\, :T_{\mu\nu}:\rangle$ of the \emph{full}  expectation value
of the stress-energy tensor $\langle\, T_{\mu\nu}\rangle$) and
therefore cannot be (directly) used to derive restrictions like
\eqref{eq:plotn}.}, though some arguments in its substantiation were
brought forward  in \cite{Qicurv}. Moreover, in their recent paper
\cite{OluGr}  Olum and Graham showed that a system of two interacting
scalar fields may violate \eqref{eq:QI}. (It is especially interesting
that the violation takes place  \emph{near a domain wall}. One might
speculate on this ground that perhaps the similarity of the shortcuts
considered in the introduction with the domains of false vacuum has
far-reaching consequences.) Thus the inequality \eqref{eq:QI} is, at
least, non-universal.

We shall, however, explore another possibility. In what follows we
demonstrate that some (perhaps most) shortcuts are not excluded even if
QI does hold.

\subsection{$\bm{\tot}$ indeterminate}
\label{sec:unk}

The derivation of \eqref{eq:est} rests heavily on the assumption that
(at least in $\Xi$) the  components of the Riemann and the Einstein
tensors are roughly of the same order: in the key relation
\begin{equation}\label{eq:EiRi}
\dis^{-4}\sim \rho^2
\end{equation}
(see the end of the chain \eqref{eq:chain}) the left hand side is
determined by the former and the right hand side by the latter.
Physically this assumption means that the case is considered in which
the exotic matter is generated mostly by  the curvature produced by
this same matter (for a moment we ignore other possible mechanisms of
generating exotic matter, such as mirrors). But the vacuum
polarization in a point is determined (among other things) by the Weyl
tensor $C_{\alpha\beta\rho\sigma}$ and not exclusively by the Ricci
tensor $R_{\alpha\beta}$.
% $$
% C_{\alpha\beta\rho\sigma}=R_{\alpha\beta\rho\sigma}
% +g_{\alpha[\sigma}R_{\rho]\beta}+g_{\beta[\rho}R_{\sigma]\alpha}
% +\frac{1}{3}Rg_{\alpha[\rho}g_{\sigma]\beta}.
% $$
 In particular, the anomalous trace of the
conformal scalar field is given \cite{BirDav} by
$$
\langle T^\mu_\mu \rangle = \frac{1}{2880\pi^2} \Bigl(
C_{\alpha\beta\rho\sigma}C^{\alpha\beta\rho\sigma}
+ R_{\alpha\beta}R^{\alpha\beta} -\frac{1}{3}R^2 +\square R \Bigr).
$$
So, one expects \eqref{eq:EiRi} to be true only when
\begin{equation}\label{eq:Cneg}
\max |C_{\alpha\beta\gamma\delta}|/ \max |R_{\alpha\beta}|\lesssim 1,
\end{equation}
a condition that breaks down more often than not. For example, in
\emph{any} curved and empty region (e.~g. in the vicinity of any star)
$$\max |C_{\alpha\beta\gamma\delta}|/ \max |R_{\alpha\beta}|=\infty.$$
And relaxing the condition \eqref{eq:Cneg} (and hence \eqref{eq:EiRi})
one immediately removes the restriction \eqref{eq:est} and invalidates
\eqref{eq:plotn}.

Another situation in which the quantum inequality does not imply
\eqref{eq:est} and \eqref{eq:plotn} concerns wormholes. The point is
that QI presumably holds for a sufficiently small
$\Delta$ because `a curved spacetime appears flat if restricted to a
sufficiently small region' \cite{Qicurv} and hence the
condition \eqref{eq:supp}. A spacetime, however, may differ from the
Minkowski space in  \emph{global} properties. As is well known,
$\varrho$ is negative and constant (in contradiction to
\eqref{eq:QI}) even in a flat spacetime, if the latter is warped into a
cylinder (the Casimir effect). To handle such situations it was
proposed in  \cite{Qicurv} to supplement \eqref{eq:supp} with a
requirement that $ \Delta$ be much less than `the proper distance from
the point $\gamma(\tau_0)$ to the boundary of the spacetime'. In
particular, if in a certain direction a condition of periodicity
--- of length $L$ --- is imposed on a field (that is, a field on a
cylinder, or, say, between mirrors is considered), then it is required
that at least\footnote{In
\cite{Qicurv} even a more restrictive requirement $\Delta\lesssim
L/\gamma$ is imposed, where $\gamma\geqslant 1$.}
\begin{equation}\label{eq:disbcyl}
  \Delta\lesssim L.
\end{equation}
This condition cannot be significantly weakened since for the massless
scalar field and a static observer we have \cite{BirDav} in the
Casimir case
\begin{equation}\label{eq:Cas}
\varrho=-\pi^2L^{-4}/1440.
\end{equation}

To see the implications of \eqref{eq:disbcyl} for shortcuts, consider
a spacetime $M_T$ obtained by the following procedure (see
figure~\ref{fig:worm}a).
\karti{t,h,b}{0.4\textwidth}{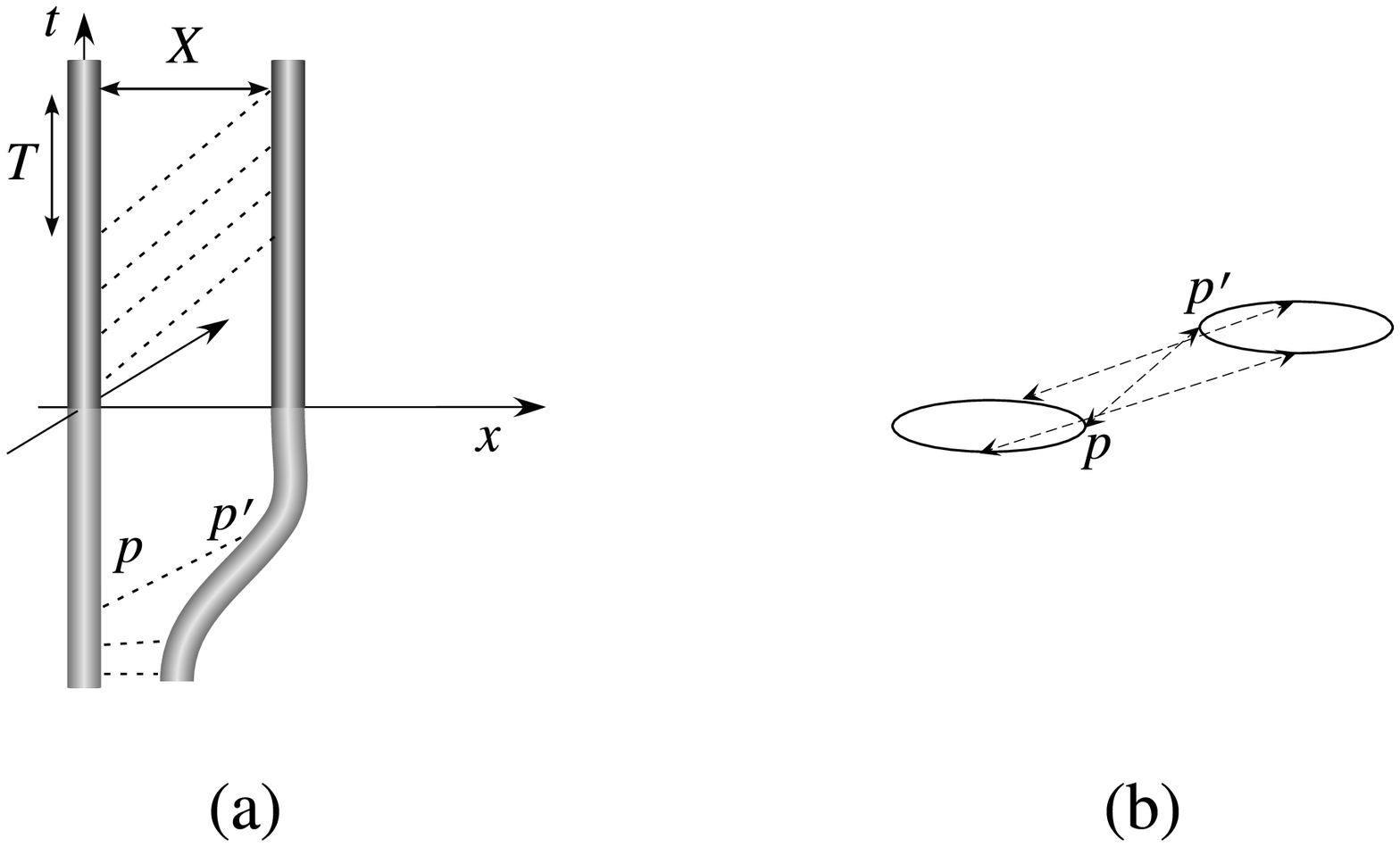}{\label{fig:worm}(a) The  throat of
the wormhole is required to be constant. So if $p$ is identified with
$p'$, then the other pairs of identified points satisfy $\tau=\tau'$,
where $\tau^({}'^)$ is the proper time measured along the cylinder
from $p^({}'^)$ to the corresponding point. Accordingly, $T$ changes
if a mouth moves. (b) The dashed lines show how the identification
must be performed.}
First, two thin vertical, i.~e.\ parallel to
the $t$-axis, cylinders are removed from the Minkowski space.
Then some (close) vicinities of the boundaries of
the holes are appropriately curved (so that the resulting spacetime be
smooth). Finally, the boundaries are identified according to the
following rule: each section $t=t_0$ of the left cylinder, which is a two
dimensional sphere, is glued to the section $t=t_0 + T$ of the right
cylinder. The identification of the
two spheres is performed so (see figure~\ref{fig:worm}b) that the
segment $(pp')$, which lies in the $(t,x)$ plane and connects the
spheres, becomes a circle.

The spacetime $M_T$ describes a wormhole whose
both mouths are at rest with respect to an inertial observer located
in the Minkowskian part of the shortcut (i.~e.\ in $U$, see the
definition of the shortcut). Given such a wormhole it is relatively
easy to impart a desired value of $T$ to it. All one needs is to move
one of the mouths \cite{MTY} without changing the geometry of the
throat (and thus without spending much energy).

Since shortcuts by definition must be globally
hyperbolic we restrict ourselves to the spaces with $X>T$ (lest the
spacetime contain closed causal curves with all the ensuing complications
\cite{MTY,par}). Among these, $M_T$ with $X\approx T$ are  most
interesting. On the one hand, such a wormhole may be an efficient
shortcut: to make a distant journey one acquires a wormhole with both
its mouths initially located near the Earth and takes one of them with
him/her. Moving  at a high  speed the traveler reaches the
destination in a short (proper) time $\Delta\tau$ (by this time $X$
becomes $\approx T$). The return trip is made through the wormhole and
(neglecting the time spent on traversing the wormhole) the traveler
returns to the Earth in $\Delta\tau$ (by the terrestrial clock) after
the departure.

And, on the other hand, such $M_T$ \emph{are free from the
restrictions (\ref{eq:est},\ref{eq:plotn}) imposed by QI.} Indeed, the
section $y=z=0$ of $M_T$ is a cylinder with $L=\sqrt{X^2-T^2}$. So,
whatever $\varrho$ is required by the Einstein equations to
support the wormhole, it suffices to make $L$ sufficiently small
(i.~e.\ $X$ sufficiently close to $L$) and the quantum inequality,
owing to \eqref{eq:disbcyl}, will be satisfied. Note that $L$, if
desired, can be made \emph{arbitrarily} small (e.~g.\ $L=l_{\rm Pl}$)
without any damage to the traversability of the wormhole: the radius
of its throat and the distance $X$ between the mouths will remain
macroscopic.
\par
Moreover, by analogy with the Casimir effect
\eqref{eq:Cas}, it is reasonable to assume that $\varrho$ in such a
wormhole will be large ($\sim L^{-4}$), which would relieve one of
having to seek additional sources of exotic matter.

\subsection{$\bm{\tot}$ large, but meaningless}
Suppose working within  classical electrostatics  we discover that for
some field configuration  the  energy $E_\Upsilon$ of the field $\bm E(x)$
contained in a region $\Upsilon$ is unphysically large. Say,
\begin{equation}\label{eq:elec}
E_\Upsilon=\frac{1}{8\pi}\int_\Upsilon E^2(x) \,\rmd^3 x = 10^{32} M_\odot.
\end{equation}
Should we conclude based on \eqref{eq:elec} \emph{alone} that the
configuration  $\bm E(x)$ is unphysical? The answer is
obviously negative. \eqref{eq:elec} is true, for example, when $\bm
E(x)$ is just the field of a pointlike charge and $\Upsilon$ is the
complement to a (small enough) ball $B_{r_c}=\{x\colon\:r(x)<r_c\}$
around the charge. The reason why such a respectable solution of the
Maxwell equations involves such huge energies  is
trivial. $\bm E(x)$ is not the \emph{real} value of the field in $x$, but
only its value
\emph{assuming} that the effects lying beyond the model
(classical electrostatics) are negligible, which  is certainly not
true when the main contribution comes from a trans-Planckian region
$r\sim r_c$. Of course the failure to predict the field, or the energy
density, in $B_{r_c}$ does not compromise electrostatics. It only
reminds us that the values of the field $\bm
E$, or potential $\varphi$ in a point, are not physically meaningful
but rather their averages $\bar{\bm E}$, $\bar \varphi$ are, defined,
say, as
\begin{equation}\label{eq:fluc}
\bar \varphi(x_0)=\frac{3}{4\pi r_l^3}\int_{B_{r_l}}\varphi \,
\rmd^3x,
\end{equation}
where $r_l$ is chosen so large that quantum fluctuations do not affect
$\bar \varphi$. While our judgments about
$\varphi(x_0)$, say, are valid only as far as it is possible to find
$r_l$ such that $\bar \varphi(x_0)-\varphi(x_0)$ is negligible.
 \par
The same considerations fully apply to semiclassical gravity. It is
not supposed to predict (correctly) the values of the relevant
quantities (such as $g_{\mu\nu}$,  \av{T_{\mu\nu}}, etc.) unless the
contribution of quantum gravitational effects can be neglected. Which
means that the enormous values of \tot\ may testify not to the fact
that the involved energies are unphysically large, but just that \tot\
is found incorrectly (that is with illegal neglect of quantum
corrections).

It is worth noting that  metrics of this kind --- on the one hand they
are solutions of semiclassical equations and on the other hand they
violate WEC in regions $\Xi$ so small that one cannot properly
assess \tot\ by means of semiclassical gravity --- arise naturally in
constructing macroscopic self-maintained wormholes. Consider, for
example, a metric \cite{MWH}
\begin{equation}
 \rmd s^2= \Omega^2(\xi)\big[
-\rmd\tau^2+\rmd\xi^2+
K^2(\xi)(\rmd\theta^2+\sin^2\theta \,\rmd\phi^2)\big],
\label{metric}
\end{equation}
where $K$ and $\Omega$ are positive even functions, such that $K(\xi)$
and $e^{-\xi/\xi_0}\Omega(\xi) $ tend to non-zero constants at large
$\xi$. This metric is increasingly flat, (that is the gravitational
field falls with $|\xi| $, though it does not vanish completely and so
\eqref{metric} is not a shortcut), which means that it describes a
wormhole\footnote{Though the metric \eqref{metric} is spherically
symmetric and static it is not of the Morris-Thorne type
\cite{Tho}. So  it  may seem somewhat surprising that it is,
nevertheless, a wormhole. It becomes obvious, however, in
appropriately chosen coordinates.}. No restrictions are imposed on
$\min{\Omega K}$, which is the radius of the wormhole's throat (again
the \emph{actual} radius, i.~e.\ the quantity that matters, when one
decides whether a wormhole can be traversed, is of course
$\min{\bar\Omega\bar K}$). In particular, it may be macroscopic
qualifying thus the wormhole as traversable. The importance of such
wormholes lies in the fact that some of them are solutions of the
Einstein equations
\eqref{eq:Ein} with $ \langle\, T_{\mu\nu}\rangle$ being the vacuum
expectation value of the stress-energy tensor of a realistic (i.~e.\
electromagnetic, neutrino, or massless scalar)  field. That is they
describe the result of the following scenario: at some moment --- say,
at the Big Bang --- a wormhole comes into existence. The non-flatness
of the spacetime near the throat polarizes the vacuum and the term $
\langle\, T_{00}\rangle$, if it is negative and large enough,
supports the wormhole and prevents it from collapsing. If it fails,
the wormhole begins to collapse and, correspondingly, the vacuum
polarization increases. The process goes on until the equilibrium is
(hopefully) found and the wormhole acquires the shape  \eqref{metric}.
How much energy this process requires is, of course, anybody's
guess.
\par Remarkably, such solutions have $\Omega$ oscillating in the
throat (i.~e.\ at $\xi\sim 0$). To leading order it may have the form
\cite{MWH}
$$\Omega_0\exp[\epsilon\sin (\xi/\xi_0)]
$$
with $\epsilon,\xi_0\ll 1$,
so the oscillations have exceedingly small magnitude and wavelength.
The energy density of the field (if found according to the
semiclassical rules) changes its sign with the sine and thus $\Xi$ is
a set of concentric spherical layers, each $\pi\xi_0\ll 1$ thick.
Therefore, as discussed above, one can hardly expect that
\tot\ defined by \eqref{eq:etot} is something measurable.
Note, however, that it is quite
different with the components of the metric: we can take $r_l$ in
\eqref{eq:fluc} to be, say, of the order of $100l_{\rm Pl}$ and check
that $\bar\Omega\approx\Omega$. Thus the model contains quantities of
two types --- some are physically meaningful and trustworthy as long
as we trust semiclassical gravity (these, for example, are
$\bar\Omega$, $\bar K$, etc.), and the others are purely auxiliary,
devoid of any specific physical meaning ($\Omega$, \tot, etc.).
\subsection{$\bm{\tot}$ small}
\subsubsection{`Portal'}
\label{subs:portal}
In this section we show that in the case of a wormhole just by
abandoning the spherical symmetry it is possible to reduce $V_\Xi$
(and thus \tot) drastically --- by 35 orders in this case. The
proposed wormhole\footnote{which is, in fact, a globally hyperbolic
analog of the `dihedral' wormhole \cite{viss}.} also has another
advantage: a traveler taking the short cut moves all the time in a
flat region and need not plunge into the Planck-density matter
\eqref{eq:est}.
\par
Consider a  spacetime $W$
\begin{equation*}
    \rmd s^2=-\rmd t^2 + 4(\varepsilon^2(\eta) + \eta^2)(\rmd \eta^2 +
    \eta^2\rmd\psi^2) + \rho^2\rmd\phi^2,
\end{equation*}
where $\varepsilon$ is a smooth even function whose support is the
region $\mathscr E\equiv \{\eta<\eta_\varepsilon\}$;
\begin{equation*}
 \rho\equiv \rho_0 - \eta^2\cos 2\psi;
\end{equation*}
and $\rho_0$, $\eta_\varepsilon$ are positive constants
$\rho_0\gg\eta^2_\varepsilon$. As usual, it is understood that
$\eta,\rho\geqslant 0$, $\phi=\phi+2\pi$, $\psi=\psi+2\pi$ and that
the points with $\eta=0$ differing only by $\psi$ are identified, as
are the points with $\rho=0$ differing only by $\phi$.

To visualize  the structure of $W$ and to check that the singularity
in $\rho=0$ is coordinate, consider, first, the region $W-\mathscr E$.
Defining $z\equiv \eta^2\sin 2\psi$ we isometrically map $W-\mathscr
E$ on a spacetime $U$, which is the Minkowski space
$$
\rmd s^2 =-\rmd t^2 +
 \rmd z^2 + \rmd \rho^2 + \rho^2\rmd\phi^2,\qquad
$$
from which at each $t$ a solid torus $\Xi=\{(\rho - \rho_0)^2 + z^2 <
\eta_\varepsilon^4\}$ is removed:
$$
(W-\mathscr E )\xrightarrow{\zeta}U\equiv\mink^4-\Xi\times\mink^1.
$$
Locally $\zeta$ is an isometry. At the same time it sends each pair of
points $(t,\phi,\eta,\psi)$, $(t,\phi,\eta,\psi+\pi)$ to a single
point $(t,\phi,\rho,z)\in U$ as shown in figure~\ref{fig:gip}a.
\karti{t,h,b}{0.4\textwidth}{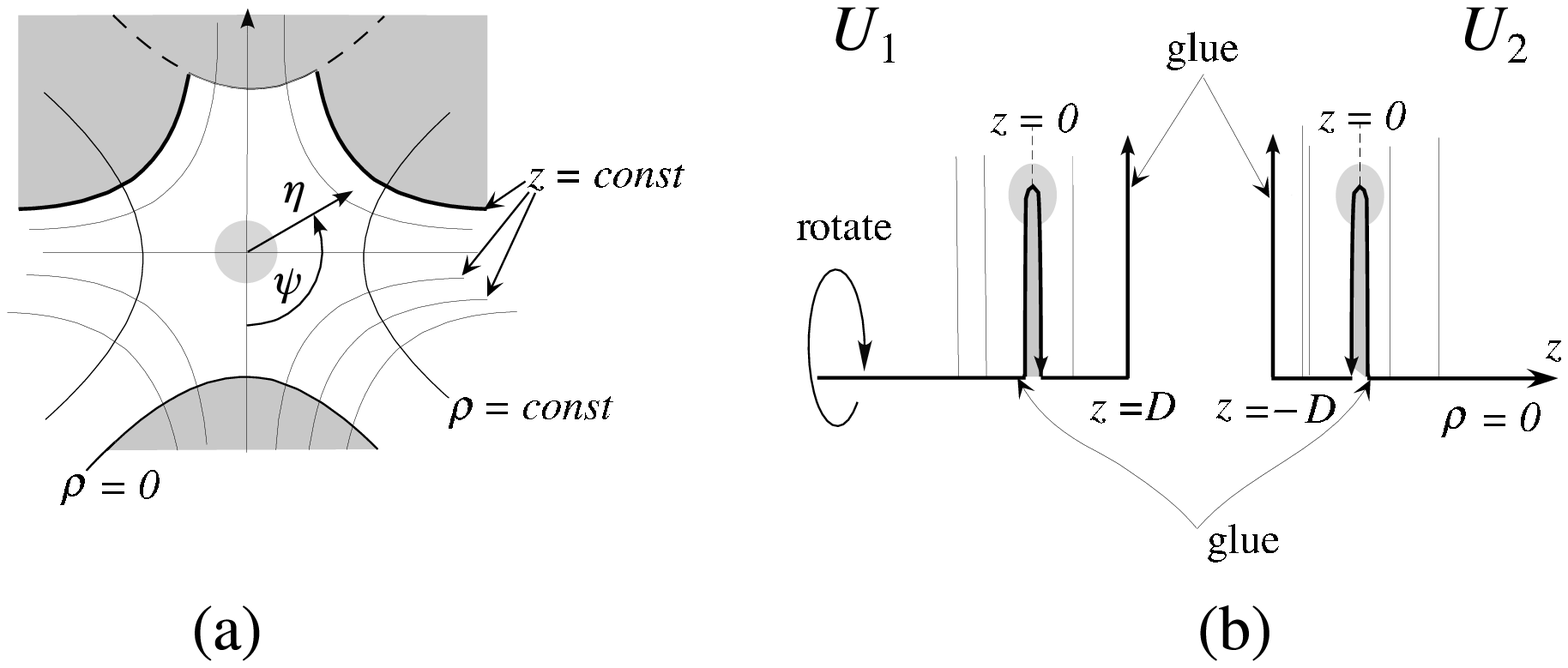}{A section $\phi=t=0$ of
the spacetime $P$. \label{fig:gip}The dark gray regions do not
belong to $P$. The thick (segments of) hyperbolae are identified.
$\zeta$ maps the left (right) half of  (a) to the left (right)
half of (b). The spacetime is flat except in the light gray
regions.}
So, $W-\mathscr E $ is the two-fold covering of $U$. To put it another
way, $W-\mathscr E$ can be constructed by taking two copies --- let us
call them $U_1$ and $U_2$ --- of $U$, cutting each of them along the
disk $(z=0,\:\rho <
\rho_0)$, and gluing the right bank of either cut to left bank of the
other.

$W$ is a wormhole connecting two `different spacetimes' $U_1$ and
$U_2$. Now we shall construct from it a shortcut $P$, i.~e., in this
instance, a wormhole that connects remote parts of the `same'
spacetime. To this end we remove the region $z>D$ ($D$ is a constant
greater than $\eta_\varepsilon^2$) from $U_1$ and the region $z<-D$
from $U_2$. $P$ is obtained by gluing together the two boundaries ---
$z=D$ in $U_1$ and  $z=-D$ in $U_2$  (note that the surgery takes
place in the Minkowski part of $W$ and thus, obviously, does not give
rise to any singularities).

Alternatively construction of $P$ can be described in terms of
$\eta$ and $\psi$. First, remove from the plane $(\eta,\psi)$ the
regions $\rho<0$ (these are the interiors of the upper and the lower
hyperbolae in figure~\ref{fig:gip}a). If we  rotate now the remaining
part of  the plane with respect to the $z$-axis, we would obtain $W$.
And $P$ is obtained if before the rotation we remove the regions
$$
\sin \psi<0,\quad \eta^2\sin 2\psi>D
\qquad\text{and}\qquad
\sin \psi>0,\quad \eta^2\sin 2\psi<-D
$$
(i.~e.\ the interiors of the left and the right hyperbolae) and
identify their boundaries.

$P$ is a globally hyperbolic and static wormhole. Its spacelike
section $P_{(3)}$ has the structure shown in
\karti{b,h,t}{0.3\textwidth}{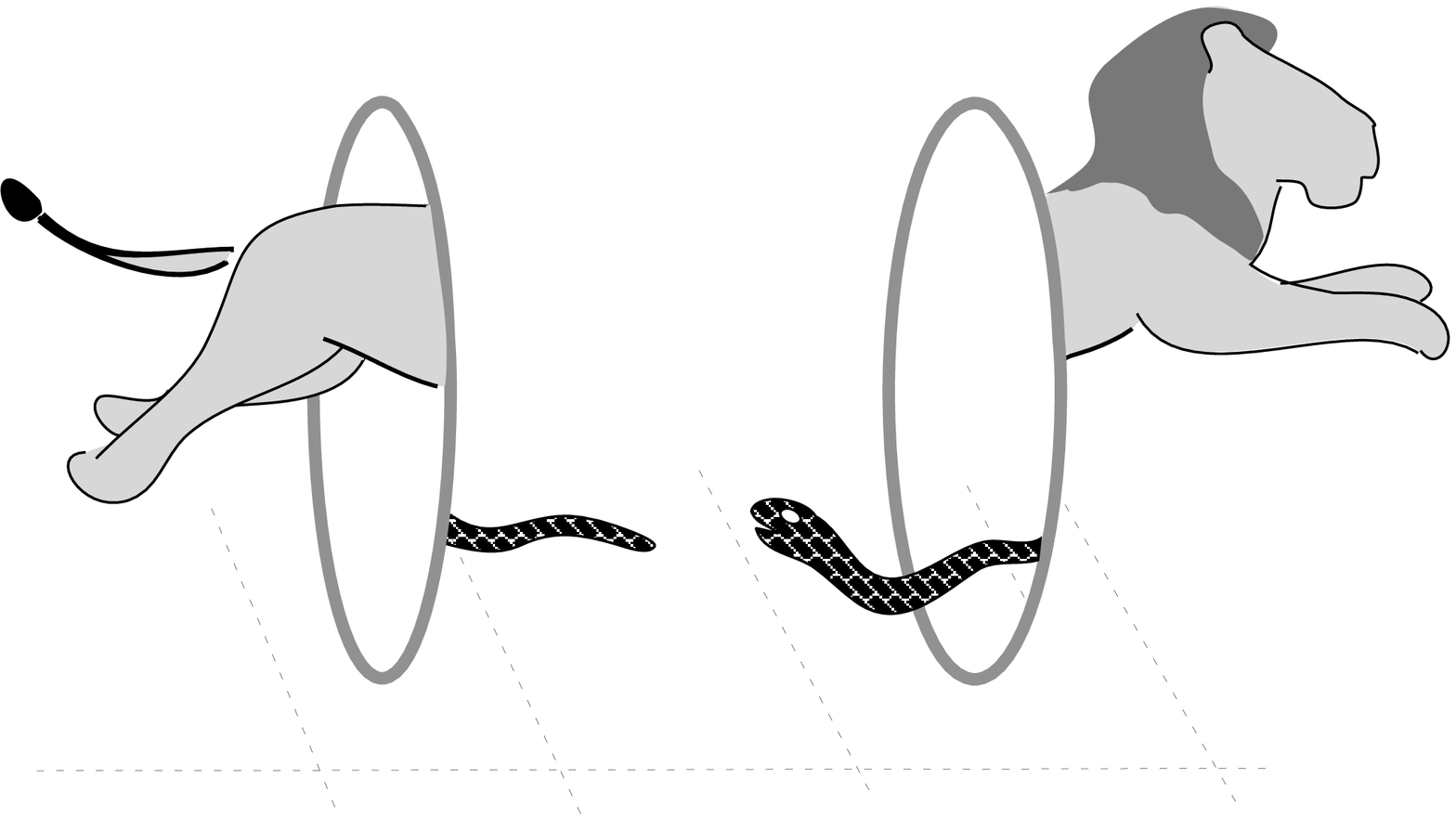}{\label{fig:portal} The
`distance between the hoops' (defined, for example, as the length of
the snake) is $2D$. The radius of each is $\rho_0$ and the thickness
is $\sim \eta_\varepsilon^2$.}
figure~\ref{fig:portal}. Outside some compact region $P_{(3)}$ is
just the Euclidean space. And inside this region the space is flat
too, except for two hoops (which, in fact, are a single hoop). A
traveler passing through one of the hoops instantly finds himself
emerging from the other one (remarkably, throughout the whole journey
the spacetime around the traveler remains empty and flat).

To estimate the required \tot\ let us choose the following
$\varepsilon$
$$
\varepsilon^2=\frac{1}{4}(\eta_\varepsilon +
\eta^2/\eta_\varepsilon)^2-\eta^2\quad \text{at }
\eta\leqslant\eta_\varepsilon,\qquad
\varepsilon=0\quad \text{at }\eta\geqslant\eta_\varepsilon,
$$
so that the metric and its
first derivatives are continuous in $\eta_\varepsilon$ (cf.
footnote~\ref{fo:smo}). With such a choice of $\varepsilon$
$$
G_{\hat t\hat t}= -\frac{4\eta_\varepsilon
^4}{(\eta^2+\eta_\varepsilon^2)^4}, \qquad
V_\Xi=\frac{14}{3}\pi^2\rho_0\eta_\varepsilon^4,
$$
whence $\tot\sim\rho_0$. So, to support a human-sized wormhole of this
type it would suffice $\tot\approx 10^{-2}M_\odot$ of exotic matter.
This trifling, in comparison with \eqref{eq:plotn}, energy is about
the energy of a supernova. QI, if it holds, does not change this
estimate in any way. As with the absurdly benign wormhole (see
section~\ref{subsec:tot}) it only requires that the hoops  be
thin (but not \emph{that} thin in this case): $\eta_\varepsilon\sim 1$.

\subsubsection{Van Den Broeck's trick}
In fact, \tot\ can be reduced further by tens of orders. Consider the
metric \eqref{eq:r(l)}, where this time $l\geqslant -l_0$, $r(-l_0)=0$
(so, the spacetime is $\rea^4$ and not a wormhole), and $r(l)$
satisfies the following conditions
\begin{equation}\label{eq:R}
\ogr{r}{l\neq -l_0}>0,\qquad \ogr{r'}{|l|>l_2}=1,\qquad
 |r'|\leqslant 1,\qquad
   r''>0\; \Leftrightarrow\; |l|<l_1,
\end{equation}
where $l_i$ are positive constants: $l_2<l_1<l_0$   (see figure
\ref{fig:pocket}a).
\karti{t,b}{0.4\textwidth}{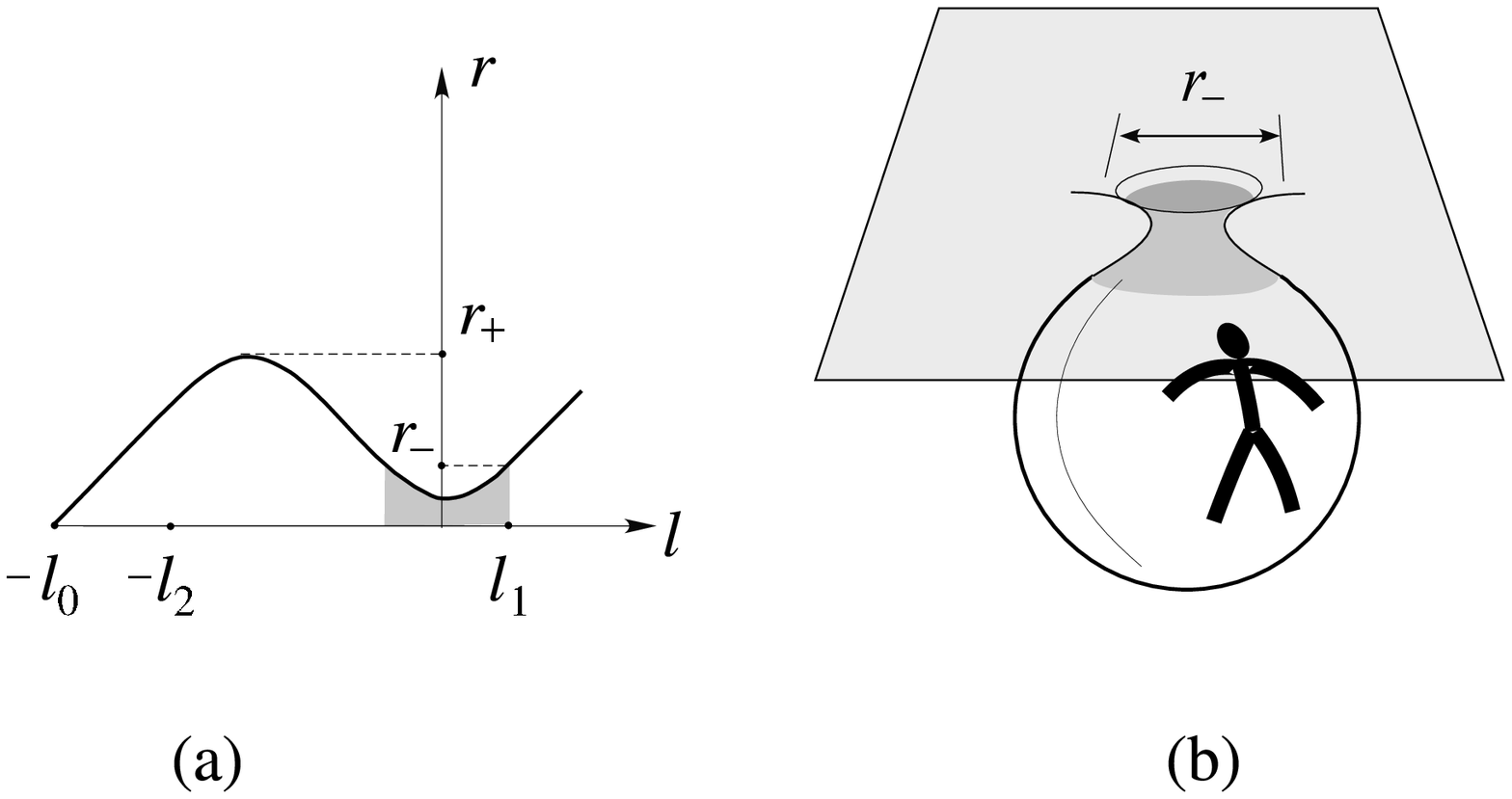}{\label{fig:pocket} With just
$\sim -10^{-5}\,$g of exotic matter one can make a Planck-size (in the
view of an outside observer) capsule be arbitrarily roomy.} The
spacetime (\ref{eq:r(l)},\ref{eq:R}) (or, rather, its section
$t,\theta=0$) is depicted in figure~\ref{fig:pocket}b. The Einstein
tensor for the metric~\eqref{eq:r(l)} can be easily found (see, e.~g.,
(14.52) of \cite{MTW}):
$$
G_{\hat t\hat t}=\frac{1-{r'}^2-2rr''}{r^2},\quad
G_{\hat t\hat t} + G_{\hat r\hat r}=-\frac{2r''}{r},\quad
G_{\hat t\hat t} + G_{\hat \nu\hat \nu}=\frac{1-{r'}^2-rr''}{r^2},
$$
where $\nu=\phi,\theta$. It follows that WEC breaks down (only)
in the spherical layer $l\in(-l_1,l_1)$.

Let us choose\footnote{We take $r$ different from that proposed in the
original paper \cite{Broeck}, because the latter, first, leads to a much
(by $\sim 10^{36}$) greater amount of exotic matter and, second,
violates the quantum inequality (e.~g.\ in $w=0.98$).}
$r=\frac{1}{2l_1}l^2+\frac{l_1}{2}$ on the interval
$l\in(-l_1,l_1)$. Then $G_{\hat t\hat t}\geqslant -8l_1^2$ and, since
the volume of the layer is $V_\Xi\sim 4\pi^2 l_1^3$, the total
amount of exotic matter is $\sim 10^2 l_1$. We can (and, if
\eqref{eq:est} holds, we must) take $l_1\sim 1$ and hence it takes
only $\sim -10^{-3}\,$g of exotic matter to support the shape of the
spacetime in discussion.

The pocket~(\ref{eq:r(l)},\ref{eq:R}) itself is not a shortcut, but
an effective means for reducing the amount of negative energy required
for a regular shortcut. The point is that, while the whole
structure is enclosed within a sphere of the radius $r_-$ from the
point of view of an outside observer (see figure~\ref{fig:pocket}b),
$r_+$ can be chosen large enough to accommodate a passenger, or
cargo. In other words, to transport a macroscopic passenger we can use
now, say, a portal described in the previous subsection with
$\rho_0=10r_-\sim 10l_{\mathrm{Pl}}$, which would require only $\sim
-10^{-4}\,$g of exotic matter (in addition to  $ -10^{-3}\,$g
spent on sustaining the pocket).
\section*{Acknowledgements}
I am grateful to  L.~H. Ford and M.~J. Pfenning for their critical
comments and to  C.~J.~Fewster for his remarks on the `difference
quantum inequality'.
\appendix*
\section{}
In this section we construct a  spherically symmetric spacetime $M$ with
the following properties (cf. subsection~\ref{subs:wec}):
\begin{enumerate}
  \item $M$ is  Schwarzschild outside a cylinder  $N=B_{r_o}\times\mink^1$;
  \item The Weak energy condition holds in the whole  $M$;
  \item The  minimal time taken by a trip through $B_{r_o}$ from a point with
  $r>r_o$ to  the diametrically opposite point decreases with time.
\end{enumerate}
First, let us introduce the following two functions
\begin{equation}\label{eq:themet}
m(r)\equiv m_0r^{-1/3}\exp\int_{r_h}^r\frac{\vartheta(x)\,\rmd
x}{3x},
\end{equation}
\begin{equation}\label{eq:defep}
\epsilon(r)\equiv \int_{r_o}^r
(r-x)\frac{x^2(m'(x)x^{-2})'}{x-2m(x)}\varphi(x) \,\rmd x,
\end{equation}
where $m_0$, and $r_h<r_o$ are some positive constants. $\vartheta$,
$\varphi$ are smooth positive functions, and the former is subject to
the following conditions
\begin{equation*}
\ogr{\vartheta}{r<r_h}=10,\quad\ogr{\vartheta}{r>r_o}=1,
\qquad \vartheta'\leqslant 0.
\end{equation*}
For later use note a few simple properties of the functions $m(r)$,
$\epsilon(r)$:
\begin{equation}\label{eq:mass}
\ogr{m(r)}{r<r_h}=m_0r_h^{-10/3}r^3,\quad
\ogr{m(r)}{r>r_o}=const
\end{equation}
and
\begin{equation}\label{eq:mproi}
m'(r)=\frac{(\vartheta-1)m(r)}{3r}\geqslant 0,\qquad
[m'(r)/r^{2}]'=\frac{m(r)}{9r^4}[3r\vartheta'+
(\vartheta-1)(\vartheta-10)]
\leqslant 0,
\end{equation}
whence, in particular,
\begin{equation}\label{eq:thprop}
\ogr{\epsilon}{r<r_h}=const ,\qquad
\ogr{\epsilon}{r>r_o}=0.
\end{equation}
We choose $m_0$ to be so small (by \eqref{eq:mass} it is always
possible) that at $r\neq 0$
\begin{equation}\label{eq:nohor}
r>2m(r)\quad\text{ and hence}\quad\epsilon'\geqslant 0
\end{equation}
(to obtain the latter inequality, differentiate \eqref{eq:defep}, use
\eqref{eq:mproi}, and note that in the region where $\epsilon\neq 0$
(i.~e.\ inside $B_{r_o}$) the upper limit of the integral
\eqref{eq:defep} is less than the lower).
\par
Consider a (smooth by \eqref{eq:mass} and \eqref{eq:nohor}) metric
\begin{equation}\label{eq:spm}
  \rmd s^2 =-e^{2\epsilon}(1-2m/r)\rmd t^2 +
 (1-2m/r)^{-1}\rmd r^2 + r^2(\rmd
  \theta^2+\sin^2\theta\rmd\phi^2).
\end{equation}
Beyond $B_{r_o}$ (`outside the cluster') $\epsilon=0$ and $m=const$,
i.~e.\ \eqref{eq:spm} becomes just the Schwarzschild metric. Let us
check that for an appropriate choice of the free function $\varphi$
the metric \eqref{eq:spm} satisfies the weak energy condition. Using
again equations (14.52) of \cite{MTW} we find
\begin{equation}\label{eq:wec0}
 G^{\hat 0\hat 0}=2r^{-2}m'\geqslant 0.
\end{equation}
Also
\begin{equation}\label{eq:wec1}
 G^{\hat 0\hat 0} +  G^{\hat 1\hat 1}=
 2\frac{r-2m}{r^2}\,\epsilon'\geqslant 0.
\end{equation}
Finally, for $i=2,3$
$$
G^{\hat 0\hat 0} +  G^{\hat \imath\hat\imath}=
(1-2m/r)\bigr[{\epsilon'}^2  +
\frac{\epsilon'}{r}\big(1-3 \frac{m'r-m}{r-2m}\big) +
\epsilon''- \frac{r^2}{r-2m}(m'/r^{2})' \bigl].
$$
The first  term in the square brackets is nonnegative and so is the
second term when $m_0$ is small enough. Hence (after twice differentiating
\eqref{eq:defep}) we obtain
\begin{equation}\label{eq:wec2}
G^{\hat 0\hat 0} +  G^{\hat \imath\hat\imath}\geqslant
(\varphi- 1)\frac{r^2}{r-2m}(m'/r^{2})'
\end{equation}
The right hand side is  nonnegative [see \eqref{eq:mproi}] when
$\varphi<1$.

Summing up, when $m_0$ is sufficiently small the metric \eqref{eq:spm}
satisfies WEC for \emph{any} $\varphi<1$. Moreover, it is easy to show
that for some interval $\sigma\subset (r_h,r_i)$ the inequalities
(\ref{eq:wec0}--\ref{eq:wec2}) are \emph{strict}. Thus WEC holds also
for a metric \eqref{eq:spm} with $\varphi(r)$ replaced by a function
$\varphi(r)-\kappa(t)\varphi_1(r)$, where $\kappa$ and $\varphi_1$
are non-negative, $\mathop{\rm supp}\varphi_1\subset \sigma$,  and
$\kappa$, $\dot\kappa$, $\ddot\kappa$ are sufficiently small. Consider
such a metric in the case when $\kappa(t)$ grows (and,
correspondingly, $\epsilon$, which by (\ref{eq:thprop},\ref{eq:nohor})
is non-positive, grows too). In this metric, if a
 curve $\gamma\subset B$ is null, then any
$\tilde\gamma$ obtained from $\gamma$ by a translation to the future
in the $t$-direction will be \emph{timelike}. In other words, the later
one starts, the less time it will take to reach the destination, even
though the metric outside $B_{r_o}$ remains Schwarzschild.

\end{document}